\documentclass[12pt]{article}
\newlength{\dinwidth}
\newlength{\dinmargin}
\usepackage{graphicx}          

\setkeys{Gin}{keepaspectratio} 

\begin{document}
\newcommand{\f}{$F_2^{\gamma}$}

\def\ap#1#2#3   {{\em Ann. Phys. (NY)} {\bf#1} (#2) #3.}   
\def\apj#1#2#3  {{\em Astrophys. J.} {\bf#1} (#2) #3.} 
\def\apjl#1#2#3 {{\em Astrophys. J. Lett.} {\bf#1} (#2) #3.}
\def\app#1#2#3  {{\em Acta. Phys. Pol.} {\bf#1} (#2) #3.}
\def\ar#1#2#3   {{\em Ann. Rev. Nucl. Part. Sci.} {\bf#1} (#2) #3.}
\def\cpc#1#2#3  {{\em Computer Phys. Comm.} {\bf#1} (#2) #3.}
\def\err#1#2#3  {{\it Erratum} {\bf#1} (#2) #3.}
\def\ib#1#2#3   {{\it ibid.} {\bf#1} (#2) #3.}
\def\jmp#1#2#3  {{\em J. Math. Phys.} {\bf#1} (#2) #3.}
\def\ijmp#1#2#3 {{\em Int. J. Mod. Phys.} {\bf#1} (#2) #3}
\def\jetp#1#2#3 {{\em JETP Lett.} {\bf#1} (#2) #3.}
\def\jpg#1#2#3  {{\em J. Phys. G.} {\bf#1} (#2) #3.}
\def\mpl#1#2#3  {{\em Mod. Phys. Lett.} {\bf#1} (#2) #3.}
\def\nat#1#2#3  {{\em Nature (London)} {\bf#1} (#2) #3.}
\def\nc#1#2#3   {{\em Nuovo Cim.} {\bf#1} (#2) #3.}
\def\nim#1#2#3  {{\em Nucl. Instr. Meth.} {\bf#1} (#2) #3.}
\def\np#1#2#3   {{\em Nucl. Phys.} {\bf#1} (#2) #3}
\def\pcps#1#2#3 {{\em Proc. Cam. Phil. Soc.} {\bf#1} (#2) #3.}
\def\pl#1#2#3   {{\em Phys. Lett.} {\bf#1} (#2) #3}
\def\prep#1#2#3 {{\em Phys. Rep.} {\bf#1} (#2) #3}
\def\prev#1#2#3 {{\em Phys. Rev.} {\bf#1} (#2) #3}
\def\prl#1#2#3  {{\em Phys. Rev. Lett.} {\bf#1} (#2) #3}
\def\prs#1#2#3  {{\em Proc. Roy. Soc.} {\bf#1} (#2) #3.}
\def\ptp#1#2#3  {{\em Prog. Th. Phys.} {\bf#1} (#2) #3.}
\def\ps#1#2#3   {{\em Physica Scripta} {\bf#1} (#2) #3.}
\def\rmp#1#2#3  {{\em Rev. Mod. Phys.} {\bf#1} (#2) #3}
\def\rpp#1#2#3  {{\em Rep. Prog. Phys.} {\bf#1} (#2) #3.}
\def\sjnp#1#2#3 {{\em Sov. J. Nucl. Phys.} {\bf#1} (#2) #3}
\def\spj#1#2#3  {{\em Sov. Phys. JEPT} {\bf#1} (#2) #3}
\def\spu#1#2#3  {{\em Sov. Phys.-Usp.} {\bf#1} (#2) #3.}
\def\zp#1#2#3   {{\em Zeit. Phys.} {\bf#1} (#2) #3}

\title{\vspace{1cm}
\bf{ Parton distributions in the photon from 
$\gamma^*\gamma$ and  $\gamma^* p$ scattering
}
\vspace{2cm}}

\author{
 {\bf H.~Abramowicz, E.~Gurvich and A.~Levy} \\ 
{\small \sl School of Physics and Astronomy}\\ {\small \sl Raymond and 
Beverly Sackler Faculty of Exact Sciences}\\
  {\small \sl Tel--Aviv University, Tel--Aviv, Israel}
}
\date{ }
\maketitle

\vspace{3cm} 
 
\begin{abstract}
  
  Leading order parton distributions in the photon are extracted from
  the existing $F_2^\gamma$ measurements and the low-$x$ proton
  structure function.  The latter is related to the photon structure
  function by assuming Gribov factorization to hold at low $x$. The
  resulting parton distributions in the photon are found to be
  consistent with the Frankfurt--Gurvich sum rule for the photon.

\end{abstract}

\vspace{-22cm}
\begin{flushright}
TAUP 2438-97 \\
July 1997 \\
\end{flushright}

\setcounter{page}{0}
\thispagestyle{empty}
\newpage  

\section{Introduction}

The notion of the photon structure function $F_2^{\gamma}$ was
introduced in analogy to the well known nucleon case. The first
measurements of \f\ became available from $e^+ e^-$ collisions which
could be interpreted as processes in which a highly virtual photon, of
virtuality $Q^2$, probes an almost real target photon, with
virtuality $P^2 \approx$ 0. 

While the proton structure function $F_2^p$ has been well measured,
$F_2^{\gamma}$ data are poor and limited mainly to the high
Bjorken--$x$ region ($x >$ 0.05)~\cite{hawar}. This limitation comes largely
because it is not easy to measure $\gamma \gamma$ interactions at high
center of mass energies, $W$.

The measured data of \f\ have been used to determine the quark
distributions in the photon~\cite{f2greview} in much the same way as
the parton densities in the proton are determined~\cite{roberts}.
However, contrary to the proton case, the gluon density in the photon
is quite badly determined since no simple momentum sum rule can be
applied. Recently, a sum rule for the virtual target photon case ($P^2
\neq$ 0) has been devised by Gurvich and Frankfurt~\cite{fg-sumrule}.
It can be extrapolated to the real photon case ($P^2$ = 0).  Most of
the presently known parameterizations of the gluon in the photon,
$xg^\gamma(x)$, violate this sum rule.

\begin{figure}[h]
\begin{center}
  \includegraphics [bb=28 161 528 650,width=\hsize,totalheight=10cm]
  {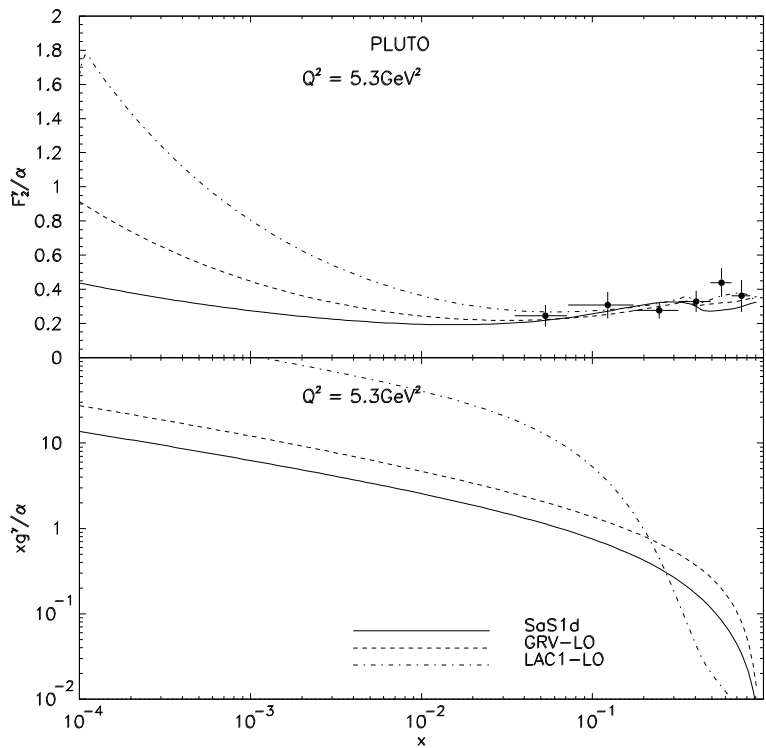}
\end{center}
\caption
{(a) Measured \f\ data at $Q^2$ = 5.3 GeV$^2$ (dots) together with
  some parton parameterizations (lines). (b) The gluon density
  distributions as obtained by the SaS (full line), GRV (dashed line)
  and LAC (dash-dotted line) parameterizations.
  }
\label{fig:f2g-5.3}
\end{figure}

As one example of the present situation concerning parton
distributions in the photon we show in figure~\ref{fig:f2g-5.3}a the
measured data of \f\ at a probing virtuality of $Q^2$ = 5.3 GeV$^2$ as
function of $x$~\cite{pluto-5.3} compared to three chosen
parameterizations SaS~\cite{ss-par}, LAC~\cite{lac} and
GRV~\cite{grv-lo}.  The data show a slight decrease as $x$ decreases.
The parameterizations, all of which give similar values in the region
where they were fitted to the data, differ appreciably in the low--$x$
region which is not constrained by the measurements. In
figure~\ref{fig:f2g-5.3}b the inferred gluon density distribution at
the same $Q^2$ is shown as function of $x$.  The distributions of the
different parameterizations differ in the whole $x$ region.

\begin{figure}[h]
\begin{center}
  \includegraphics [bb=26 140 549 669,width=\hsize,totalheight=10cm]
  {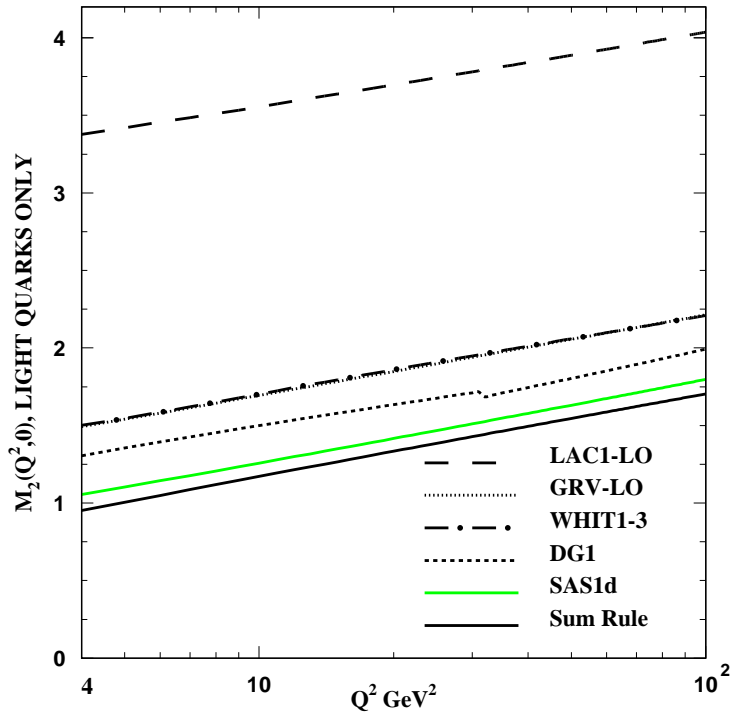}
\end{center}
\caption
 {
   The constrain on the moment $M_2(Q^2,0)$ obtained from the
Frankfurt--Gurvich sum rule for the photon (full line) as
   function of $Q^2$ for the real photon case, compared to the results
   obtained from various parton parameterizations, as denoted in the
   figure (see text).
  }
\label{fig:sumrule}
\end{figure}

In figure~\ref{fig:sumrule} we present the quantity $M_2(Q^2,0)$ which
is constrained by the Frankfurt--Gurvich sum rule~\cite{fg-sumrule} in
the real photon case, as function of $Q^2$, for the case of light
quarks only. In addition the moment is compared to the results
obtained from various parton parameterizations: WHIT1--3~\cite{white},
DG1~\cite{dg}, GRV--LO, SAS1d and LAC.  As can be seen, only the SaS
parameterization is close to the expectations of the sum rule.

In the present note we use Gribov factorization~\cite{gribov-fact} as
suggested in~\cite{al-gribov} to constrain the behaviour of \f\ at low
$x$ from the data of $F_2^p$.  We then use this extended set of data
to extract the leading order parton distributions in the photon.

\section{Gribov factorization}

Gribov factorization~\cite{gribov-fact} is based on the assumption
that at high energies the total cross section of two interacting
particles is determined by the property of the universal pomeron
trajectory. This implies relations between total cross sections of
various interacting particles.  In particular, the Gribov factorization
can be used~\cite{gribov-fact,rosner} to relate the total $\gamma
\gamma$ cross section, $\sigma_{\gamma \gamma}$, with that of
photoproduction, $\sigma_{\gamma p}$, and that of $p p$, $\sigma_{p
  p}$, all at the same center of mass energy squared $W^2$.  Using the
assumption made in~\cite{al-gribov} that at low--$x$ Gribov
factorization is applicable also for virtual photons, one can relate
the proton and the real photon structure functions in a simple
way~\cite{al-gribov}
 
\begin{equation}
F_2^\gamma(x,Q^2) = F_2^p(x,Q^2) \frac{\sigma_{\gamma p}(W^2)}
{\sigma_{p p}(W^2)}.
\label{eq:fac-f2}
\end{equation}

Relation~(\ref{eq:fac-f2}) allows the use of well measured quantities
like total cross sections and the proton structure function $F_2^p$ to
predict the values of the photon structure function \f\ in the region of
low--$x$ where equation~(\ref{eq:fac-f2}) is expected to be valid.
Since this is also the region where direct measurements of the photon
structure function are difficult and not available, the use
of~(\ref{eq:fac-f2}) provides a way to `obtain' \f\ `data' and use
them as an additional source, on top of the direct measurements of \f\ , 
to constrain the parton distributions in the photon.

\section{Results}

We have used data on \f\ in order to obtain the parton distributions
in the photon, in a similar way to the procedure used in~\cite{lac}. We
have applied a leading order evolution equation, using four flavours,
and assumed the following simple parameterization forms for the
partons, at a starting scale of $Q^2_0$ = 4 GeV$^2$:
\begin{eqnarray}
xq(x) &=& K_{sup} A e_q^2 (1+Bx) x^C 
\label{eq:param1} \\
xg(x) &=& A_g x^{B_g} (1-x)^{C_g}
\label{eq:param2}
\end{eqnarray}
where $K_{sup}$ is a suppression factor, having the value 1 for $u$ and
$d$ quarks, and $K_{sup}$ = 0.3 for $s$ and $c$ quarks. The charge of
the quarks is denoted by $e_q$ and $A, B, C, A_g, B_g$ and $C_g$ are
parameters to be determined from a fit to the \f\ data.

We have used all available \f\ data~\cite{pluto2}--~\cite{opal} 
 together with the indirect
`data' obtained through relation~(\ref{eq:fac-f2}). The latter was
obtained by using the data of the proton structure
function~\cite{hawar} for $x \leq$ 0.01. For the total photoproduction
cross section and the total $p p$ cross sections, we used the
Donnachie--Landshoff parameterizations~\cite{dl}, which give a good
representation of the total cross section data. For the measured \f\ data, 
we used as errors the statistical and systematical errors added in quadrature.
For the `data' as obtained through relation~(\ref{eq:fac-f2}) we used an 
additional systematic error of 3\% 
as an estimate of the uncertainty coming 
from assuming the Gribov factorization to hold also for the virtual
photon case at low--$x$.

\begin{figure}[h]
\begin{center}
  \includegraphics [bb=25 164 528 650,width=\hsize,totalheight=12cm]
  {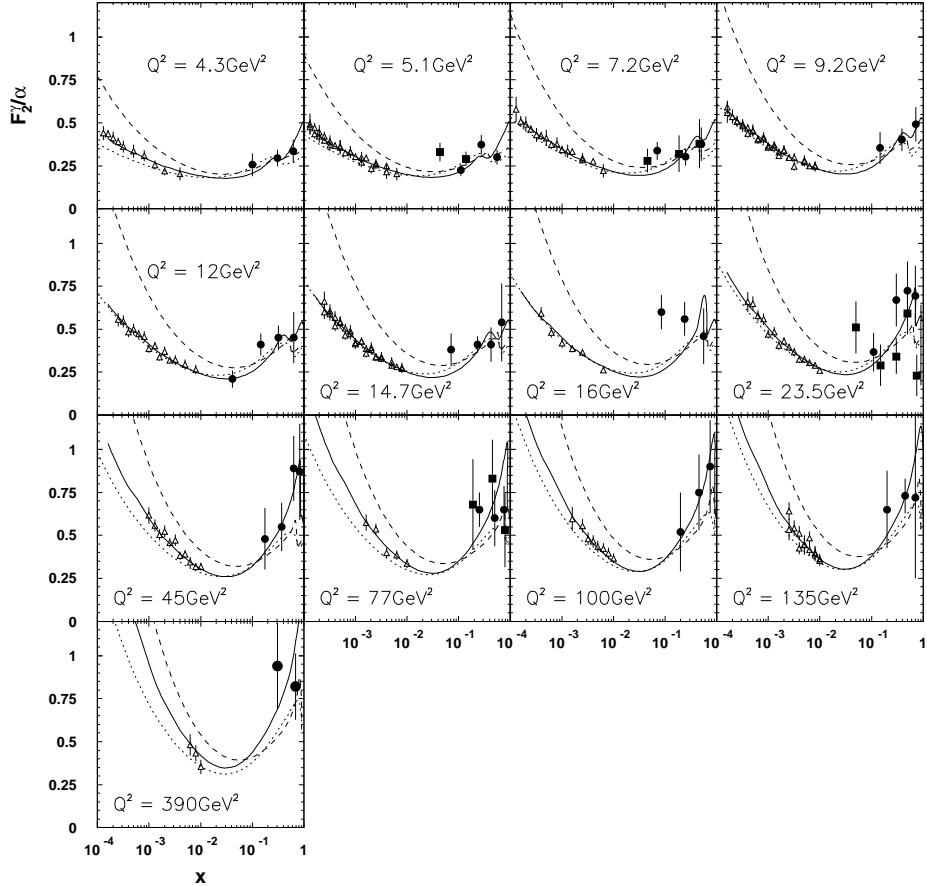}
\end{center}
\caption
{ 
  The photon structure function as function of $x$ for fixed $Q^2$
  values as indicated in the figure. The full points are the direct
  measurements and the open triangles are those obtained from $F_2^p$
  through the Gribov factorization relation (1). 
The full line is the result
  of the present fit as described in the text, the dashed line is that
  of the GRV--LO parameterization, and the dotted line is that of SaS.
}
\label{fig:f2g-all}
\end{figure}

The parameters in~(\ref{eq:param1}) and~(\ref{eq:param2})
were determined from a best fit to the data. We used 241 data points
and obtained a $\chi^2$ value of 162. The results of the best fit
are displayed in figure~\ref{fig:f2g-all}.
The directly measured photon structure function data appear as full
points, while the data obtained through the use of the Gribov
factorization are displayed as open triangles. We displayed the \f\
data of $Q^2$ = 5.09 (TPC/2$\gamma$) and 5.1 (TOPAZ) in the same
figure labeled 5.2 GeV$^2$. We also show the data of $Q^2$ of 23
(TASSO) and 24 (JADE) GeV$^2$ in one bin of $Q^2$ = 23.5 GeV$^2$. All
the low--$x$ data coming from the proton structure function have been
scaled to the value of $Q^2$ which is indicated in the figure.  The
solid curves are the results of the present parameterization. For
comparison we show as dashed lines the results of the GRV-LO
parameterization which differ substantially from the present
parameterization in the low--$x$ region. The results of the SaS
parameterization, also shown, are very close to those of the present
parameterization.

The photon structure function data exhibit a minimum in the range
$10^{-2}<x<10^{-1}$, which is broad at lower $Q^2$ and gets sharper as
$Q^2$ increases. The increase of \f\ with decreasing $x$ is not
surprising since it is an outcome of the factorization assumption
which uses the proton structure function data.

The results of the new parameterization, which give a good fit to the
data in the whole $x$ region, can be confronted with the expectations
of the Frankfurt--Gurvich sum rule. This constraint was not used in
the fit. The sum rule
using the new parameterization agree with the theoretical expectations
within 5\%. The only other parameterization which is close to fulfilling
the expected sum rule is that of SaS.

\begin{figure}[h]
\begin{center}
  \includegraphics [bb=25 164 528 650,width=\hsize,totalheight=8cm]
  {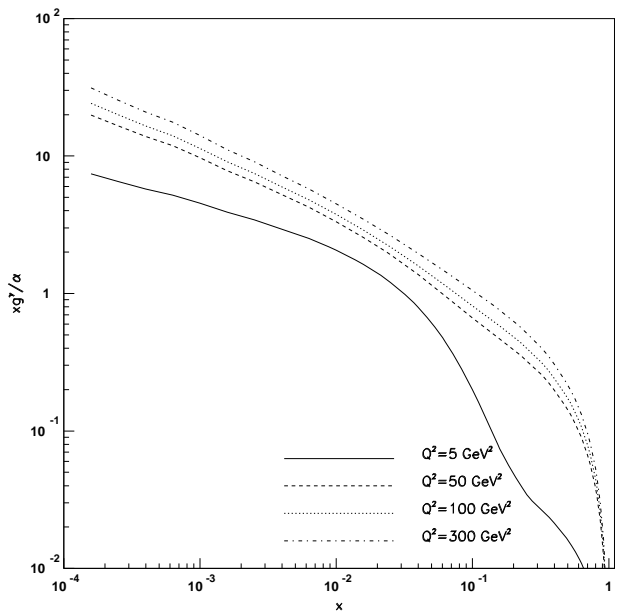}
\end{center}
\caption
{
The gluon momentum distribution in the photon as function of $x$, for 
fixed values of $Q^2$, as calculated from the present parameterization.
}
\label{fig:gluon}
\end{figure}

In figure~\ref{fig:gluon} we present the gluon momentum density as
calculated from the form~(\ref{eq:param2}) as function of $x$, for a
few fixed values of $Q^2$. The gluon density increases as $x$
decreases. One sees also the increase of the gluon momentum density
with $Q^2$, at a given value of $x$. This is a reflection of the
feature that the photon structure function has a positive scaling
violation at all values of $x$, contrary to the proton case.

\section{Conclusions}

We have extracted the leading order parton distributions in the photon
by using the existing \f\ measurements and the low--$x$ proton
structure function. The latter is related to the photon structure
function by assuming Gribov factorization to hold at low $x$. The
resulting parton distributions in the photon fulfil the
Frankfurt--Gurvich sum rule for the photon.

\section*{Acknowledgments}
 
It is a pleasure to acknowledge fruitful discussions with Prof. Lonya
Frankfurt.

\noindent This work was partially supported by the German--Israel 
Foundation (GIF).


\end{document}